\documentclass[5p,preprint]{elsarticle}

\RequirePackage{fix-cm}
\usepackage{booktabs}
\usepackage{amsmath}
\usepackage{mathptmx}   
\usepackage{graphicx}
\usepackage[misc]{ifsym}
\usepackage{array}
\usepackage[normalem]{ulem}
\usepackage{scalerel}

\newif\ifdraft
\newcommand{\draft}[1]{\ifdraft \emph{#1} \fi}
\draftfalse
\usepackage{array}
\newcolumntype{L}[1]{>{\raggedright\let\newline\\\arraybackslash\hspace{0pt}}m{#1}}
\newcolumntype{C}[1]{>{\centering\let\newline\\\arraybackslash\hspace{0pt}}m{#1}}
\newcolumntype{R}[1]{>{\raggedleft\let\newline\\\arraybackslash\hspace{0pt}}m{#1}}
\newcommand{\eg}{e.\,g.}
\newcommand{\ie}{i.\,e.}

\begin{document}

\begin{frontmatter}

\title{Predictive Performance Modeling for Distributed Computing using\\Black-Box Monitoring and Machine Learning}

\author{Carl~Witt\corref{cor1}\fnref{orcid}}
\ead{wittcarl@informatik.hu-berlin.de}
\author{Marc Bux\corref{cor2}}
\ead{buxmarcn@informatik.hu-berlin.de}
\author{Wladislaw~Gusew}
\ead{gusewwly@informatik.hu-berlin.de}
\author{Ulf~Leser}
\ead{leser@informatik.hu-berlin.de}
\address{Humboldt-Universit\"at zu Berlin, 6, Unter den Linden, Berlin, GER.}

\cortext[cor1]{Corresponding author}
\fntext[orcid]{ORCiD: 0000-0002-7918-605X}

\begin{abstract}
In many domains, the previous decade was characterized by increasing data volumes and growing complexity of computational workloads, creating new demands for highly data-parallel computing in distributed systems. 
Effective operation of these systems is challenging when facing uncertainties about the performance of jobs and tasks under varying resource configurations, \eg, for scheduling and resource allocation. 
We survey predictive performance modeling (PPM) approaches to estimate performance metrics such as execution duration, required memory or wait times of future jobs and tasks based on past performance observations.
We focus on non-intrusive methods, \ie, methods that can be applied to any workload without modification, since the workload is usually a black-box from the perspective of the systems managing the computational infrastructure.
We classify and compare sources of performance variation, predicted performance metrics, required training data, use cases, and the underlying prediction techniques. We conclude by identifying several open problems and pressing research needs in the field.
\end{abstract}

\begin{keyword}
Distributed Computing \sep Resource Management \sep Machine Learning \sep Black-Box Monitoring 
\end{keyword}

\end{frontmatter}


\section{Introduction}

Growing computational demands in science and industry are strong drivers of innovations in distributed computing, initiating, for instance, the development of grid infrastructures or cloud computing~\cite{2009arXiv0901.0131F,Hussain:2013dw}. One of the fundamental problems occurring when performing complex computations in a distributed environment is scheduling: Given a set of resources (compute nodes, network, storage) and a workload, a scheduler has to decide when and which resources (nodes) are to be assigned to which units of work (jobs or tasks). Scheduling decisions are typically guided by a goal, such as minimization of runtime or of resource usage, and often have to meet additional constraints, such as resource limits of the individual nodes or data dependencies between tasks~\cite{Casavant:1988ii}. To achieve its given goals as good as possible under the specified constraints, a scheduler requires precise estimates about the expected performance of a given job or task for a given input on given resources. Only with such knowledge it is possible to take decisions which consider their future implications, which is a pre-requisite for achieving near-optimal schedules~\cite{Casanova:2000iw}.

Unfortunately, such estimates are very difficult to obtain in practice~\cite{Matsunaga:2010kr,Smith:2004gh,Iverson:1999cu,Devarakonda:1989wl}. In most systems, the scheduler has no knowledge about the particular computation performed by a job or task, i.e., it treats everything as black-box. Accordingly, approaches involving analysis of source codes~\cite{Nudd:2000gd,Taylor:2003fo} or hand-crafted analytical models for specific operators or workloads~\cite{Khan:2016bk,Ganapathi:2009jo} are not applicable. Instead, the only available type of information are usually a few observations of performance for varying problem sizes and resource configurations extracted from log files of past executions of the same or a similar workload. The challenge is to derive precise predictions from these sparse samples. 

Due to these difficulties, many practical schedulers completely disregard such estimations~\cite{Tsafrir:2007hg}, creating schedules that potentially are far from optimal~\cite{Agullo:2016dh,Tsafrir:2007hg,Gaussier:2015fu}. Other systems delegate the task of estimating resource consumption to the user, demanding a deep understanding of the workload, the input data, and the infrastructure at hand, often leading to inaccurate estimates~\cite{Tang:2010if,RamirezAlcaraz:2011jy}.

In this paper, we survey predictive performance modeling (PPM) approaches for execution durations, resource consumption, and wait times for black-box jobs and tasks based on historical data. 
Such approaches extrapolate historical observations to future situations without requiring detailed knowledge about the internals of the workload. 
Their basic idea is to model relationships between workload and execution environment based on historical performance observations, using, for instance, regression~\cite{Harrell:2015vh} or time series analysis~\cite{Wei:2006wh}. 

Research on PPM dates back at least to the late eighties, motivated by load balancing in a distributed system~\cite{Devarakonda:1989wl}. Later, cluster computing constituted the focus of the research, followed by a series of relevant works in the context of grid computing~\cite{2009arXiv0901.0131F}. The development to this point was characterized by an increasing consideration of system heterogeneity, moving from single-core systems over clusters of homogeneous machines to data centers encompassing a battery of different node types. System heterogeneity was further increased through the recent introduction of cloud computing, where resources are typically virtualized and available in various configurations.
On the workload side, big data workloads~\cite{Dean:2008ua,Gandomi:2015hh,Hashem:2015jm} and complex analysis pipelines~\cite{Venkataraman:2016ww,Liew:2017fc} added to the diversity of the processing tasks, increasing the demand for other resources than CPU, like disk I/O, and the complexity of the resource usage patterns.

We are aware of only a few previous surveys in related fields. \cite{Pllana:2008hv,Seneviratne:2014fd} reviewed performance modeling, but focused on analytical models, breaking the black-box assumption. \cite{Jennings:2015ht,Qureshi:2014gq} survey the general field of distributed resource management but without special attention to PPM. \cite{Hutter:2014ci} surveyed machine learning methods for predicting the performance of algorithms but only considered isolated programs on a single machine, whereas we target methods for complex workloads on a distributed environment. \cite{Osman:2012cm} reviews techniques to model the performance of relational queries using queuing theory, but focuses on relational, transactional workloads.

This survey is structured as follows. 
Section~\ref{sec-distributed-computing} provides background on predictive resource management in a distributed system. 
Section~\ref{sec-taxonomy} describes our comparison scheme for PPM approaches. Section~\ref{sec-task-models} presents approaches to forecast the behavior of a single task on a single node. In Section~\ref{sec-job-models}, we review performance models for complex workloads (jobs) on a set of nodes. In Section~\ref{sec-performance-metrics} we review techniques for black-box performance monitoring. We conclude in Section~\ref{sec-open-issues} with open questions and possible research directions.

\section{Background and Scope}
\label{sec-distributed-computing}

\begin{figure}[t]
\label{fig-resources-workload}
\includegraphics[width=.48\textwidth]{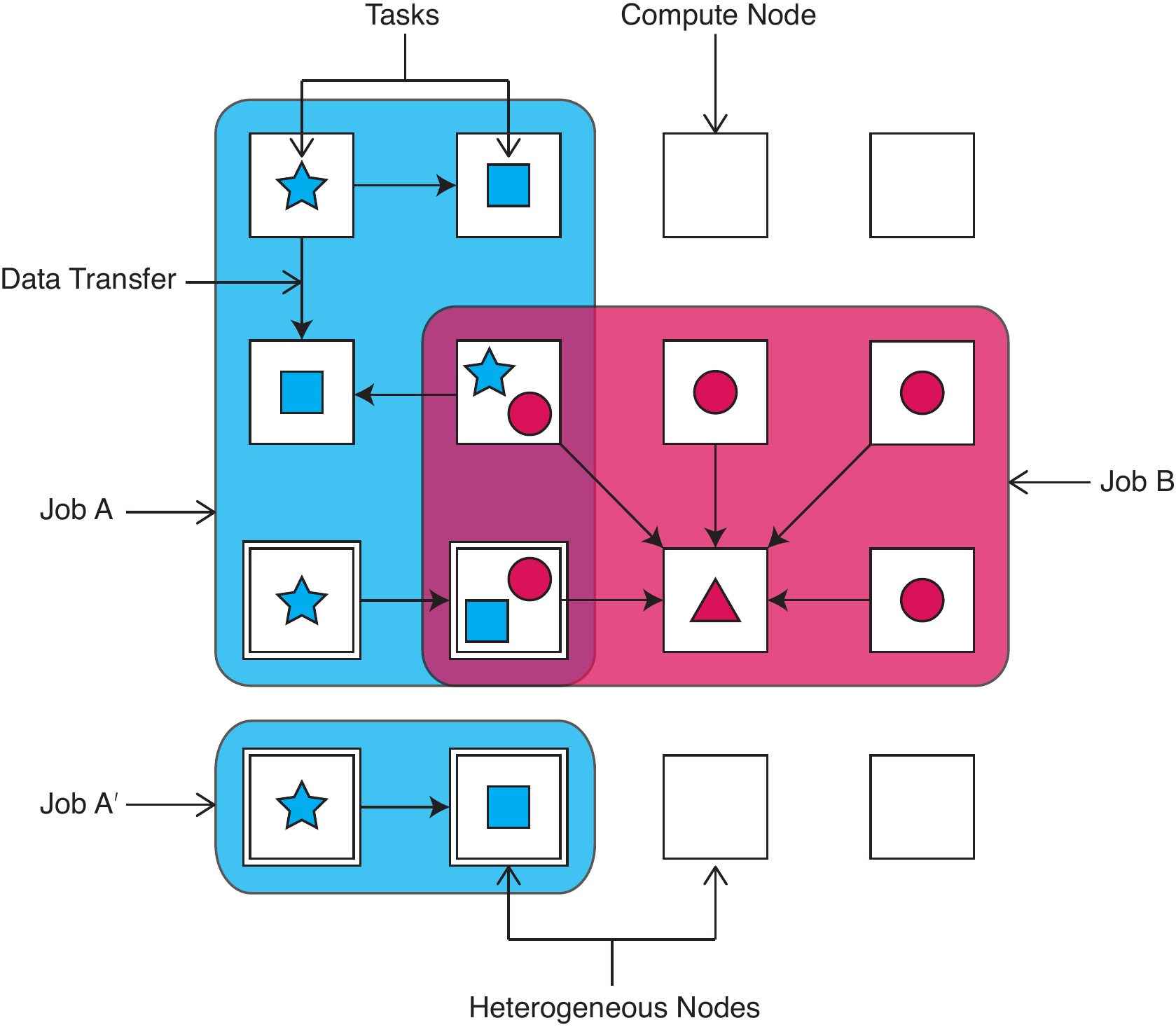}
\caption{Overview of the distributed resource and workload model. Jobs have different resource usage patterns (computation and communication), occur at different scales (Job A and Job A'), utilize compute nodes with heterogeneous hardware specifications, and experience resource contention on a shared node (\eg, tasks \protect\includegraphics[height=6pt]{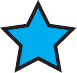} and \protect\includegraphics[height=6pt]{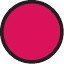}).}
\end{figure}

In this work, we survey methods for predictive performance modeling (PPM) to improve resource efficiency and user satisfaction in distributed systems. The predictions are provided either to a system component such as the scheduler or a user facing a similar decisions, \eg, the amount of resources to request for a job.
In both cases, decision making is driven by certain constraints and goals, such as fairness, deadlines, or costs.
To make valid decisions under a predefined optimization goal, PPM can be used to compare alternative resource allocations, job placement, starting time, resource types, etc. 

In the following subsections, we first give an abstract model for the most important resources within a distributed execution environment and for a workload. We then outline exemplary scheduling use cases and their information needs.

\subsection{Abstract View on Resources and Workloads}


We model a distributed computing system as a set of \emph{compute nodes} that are connected via a network. Each node encompasses various devices, such as CPUs, memory, disks, and network controllers. 
The nodes that comprise a system can vary in their available resources, featuring, for instance, different numbers of cores or different sizes of main memory.
The network may also connect nodes at varying speeds, \eg, across racks in a cluster or across sites in a grid. Note that many modern system architectures virtualize resources to improve maintainability and resource usage~\cite{Barham:2003jw}. 

Users submit \emph{jobs} to the system for execution. A job works on some input data and may have parameters, which may affect the amount of work to be done and its resource usage patterns. Some jobs can be broken down to \emph{tasks} which represent the smallest unit of work for the system. Each task is executed on a single node, but nevertheless may employ parallelism by utilizing multiple threads. 

Distributed systems, and thus their resources, are usually shared by different users. Different policies for sharing resources exist. In a system that is \emph{space-shared}, a set of resources is assigned exclusively to one job, and this assignment remains until the job finishes. In a space-shared system, jobs are typically submitted to one or more queues, implementing system properties like fairness or job priorities~\cite{Hovestadt:2003gy}. In a \emph{time-shared} system, multiple jobs can run on overlapping resources at the same time. This implies that they potentially compete for system resources, giving rise to resource contention~\cite{Zhu:2003us}. However, from a scheduling point of view, these policies are not fundamentally different, as in both cases the amount of resources and the placement need to be decided such that some optimization goal is achieved.

Our model of distributed computing covers various scenarios, including cluster, grid, and cloud computing, high performance computing and big data workloads. A very generic model is sufficient because of the versatility of the black-box approach. Although these paradigms differ strongly in their organizational aspects such as resource ownership, pricing, access, etc. black-box modeling techniques are applicable to all of them because they rely on patterns in the observed performance data rather than models of the internals of a system or workload.

\subsection{Scheduling Use Cases}
\label{sec-scheduling-use-cases}


In this section, we briefly review some of the use cases that arise when using PPM to support a scheduler in a distributed system, which is a frequent motivation for research in this field.
A scheduler is the component of a distributed system which assigns resources to jobs and tasks in a way that aims at optimizing some criterion, such as throughput, time-to-finish, energy consumption, or even monetary cost, \eg, when renting resources in a cloud. 

Scheduling heuristics exist for various scenarios, such as running independent batch jobs by various users on a cluster\cite{Delimitrou:2014hz,Tsafrir:2007hg}, or for executing workflows of interdependent tasks~\cite{FerreiradaSilva:2017gi,Topcuoglu:2002bz}.
Except for extremely constrained scenarios, scheduling problems are difficult to solve and usually NP-hard~\cite{Lenstra:1977kx}. Accordingly, there exists a rich literature on heuristics and strategies for scheduling~\cite{Dong:2006to, Smanchat:2015ex,Zhan:2015ie, Lopes:2016dp}. The focus of our survey are methods to obtain performance predictions that support informed decision making. Such performance estimates are fundamental for many approaches to scheduling; in the following, we describe, for illustration, three popular techniques all relying on accurate cost estimates.

\draft{backfilling}
Backfilling is a scheduling technique to improve the utilization and response time of a system~\cite{Mualem:2001jf}. In a space-shared compute cluster, jobs are often scheduled according to a first-come-first-served policy. A backfilling policy allows short jobs to take advantage of resources left idle by larger jobs. To make sure that a job really fits a gap between reservations, predictions of job execution duration are required (\eg ~\cite{Tsafrir:2007hg}).

\draft{DAG scheduling}
Task graph scheduling is the problem of executing computational tasks in compliance with their control and data dependencies~\cite{Kwok:1999dw}. The problem again is NP-hard, which has led to the development of numerous scheduling heuristics, like the popular Heterogeneous Earliest Finish Time heuristic~\cite{Topcuoglu:2002bz}. These heuristics typically rely on estimated execution and file transfer times~\cite{Blythe:2005eb}.

\draft{site selection}
Site selection denotes the problem of choosing, for a given job, from various eligible sites the one that best fits the scheduling goal, depending on properties like required data transfer times, expected execution times, or the current load on the sites~\cite{Li:2004iv}. Site selection was intensively studied in grid computing, motivated by the availability of various clusters for a given job~\cite{Li:2007by}. Another application area are systems that chose the best from a set of available clouds~\cite{Cunha:2017in}. Clearly, the expected runtime of a job on a site (or cloud) is an important parameter in any such decision method.

\section{Comparing Black-Box Performance Prediction Methods}
\label{sec-taxonomy}


In this section, we describe the four distinction criteria along which we compare methods for PPM: (1) Methods can be distinguished by the unit of prediction: Some predict the performance of single tasks, while others predict performance of entire jobs. (2) Another distinction lies in the metrics which are predicted, such as wall-clock time or peak memory usage. (3) Different methods take different properties of the system and tasks into consideration, i.e., they differ in the principal performance factors which their prediction model covers. (4) Approaches can be classified according to the prediction method they employ, such as classification or regression. In the following, we explain each of the criteria in more detail.

\subsection{Workload Granularity}

We divide approaches in two categories depending on the target of analysis. The first category comprises task models (Section~\ref{sec-task-models}), \ie, methods that predict the performance of a single node executing a single task.
Approaches in this category may rely on various performance information sources, such as hardware performance counters, the operating system, and binary instrumentation. 
In the context of distributed computing, the relevance of task performance prediction arises in various scenarios, \eg, as input to scheduling heuristics for several tasks with dependencies~\cite{Topcuoglu:2002bz,Liu:2015iy}. 

The second category of approaches comprises job models (Section~\ref{sec-job-models}), which assess the collective performance of a set of nodes working together to solve a workload of tasks. A standard use case for job-level performance prediction is batch scheduling in a high-performance computing system. Such systems are typically based on per-job information from previous execution, like the total amount of the requested resources or the overall execution duration.

\subsection{Performance Metrics}
\label{sec-performance-metrics}

\begin{figure}[t]
\centering
\includegraphics[width=.32\textwidth]{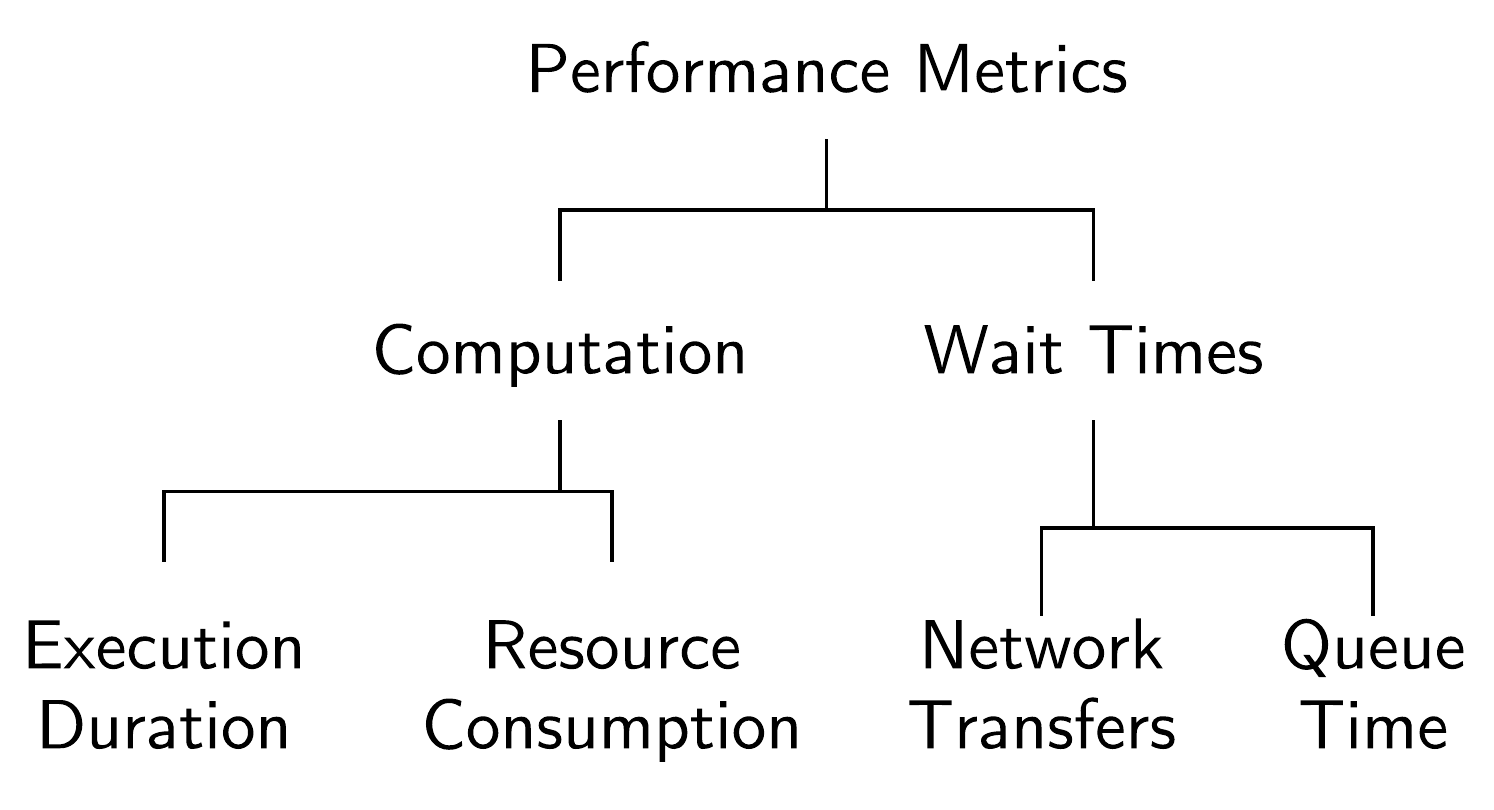}
\caption{The performance metrics capturing the different notions of performance in a distributed system.}
\label{fig-performance-metrics}
\end{figure}

\draft{section overview}
Methods for PPM differ in the metric of performance they try to assess, i.e., the concrete aspect of performance predicted. We categorize these into metrics related to computation and metrics related to wait times (see Figure~\ref{fig-performance-metrics}).

\draft{computation}
The most commonly predicted metrics are related to computation, a term which itself subsumes different aspects. The most prominent one is \emph{execution duration} (wall-clock time), \ie, the time that elapses between the beginning and the completion of a task or a job. It is typically correlated to other metrics, such as input size~\cite{Iverson:1999cu}, program arguments \cite{Lee:2007tv}, or hardware characteristics~\cite{Pfeiffer:2008jy}. Estimated wall-clock times constitute important information for schedulers, for instance to estimate at what point-in-time a given node becomes free again, or to estimate the critical path through the tasks of a job ~\cite{Kwok:1999dw}. A second perspective on computation is the \emph{amount of resources} required to complete a task or job. For instance, predicting the number of CPU instructions can be used as an intermediate metric to estimate wall-clock time~ \cite{Kuperberg:2008fr}. Other important aspects of computation are the peak amount of memory used and the amount of disk I/O or network traffic caused. Both are critical measures for a scheduler to prevent spilling from main memory to disk and contention of processes competing for constrained data exchange channels~\cite{Matsunaga:2010kr,daSilva:2015kv}. 

\draft{wait times}
A second important category of performance metrics covers \emph{wait times}, resulting from activities supplementary to the actual computing. Especially the \emph{duration of file transfers} during a task or job execution has been subject of various works. Transfer times depend mainly on the amount of data to transfer, the bandwidth of the communication channel, the current load on this channel, and the possibility to use multiple channels in parallel~\cite{Vazhkudai:2003jn, Arslan:2016up, Faerman:1999iq}. Thus, transfer times largely depend on properties of a system which are not under the exclusive control of the scheduler, making analytical models difficult to employ; instead, predictions are usually based on real-time network probing~\cite{Wolski:1999hw,Swany:2002cs,Faerman:1999iq,Vazhkudai:2003jn}. Transfer time predictions are important to estimate when all necessary data for a scheduled task will be available, or to find the fastest route to data in case multiple replica exist, or to choose a node for executing a task which has the fastest access to the data this task requires. Specific research has been devoted to the communication costs of programs written in certain paradigms, \eg, Message Passing Interface~\cite{Gropp:1999vv} or MapReduce~\cite{Dean:2008ua}.
We do not cover approaches specific to these paradigms, as the focus of this survey are black-box approaches.
Finally, considerable wait times often occur when a scheduled job has to wait for resources to become free. This effect is usually captured by measuring the \emph{time spent in a queue} waiting for resources~\cite{Downey:1997dk}. A scheduler can utilize predicted queue times to select among multiple available computation sites.

\subsection{Principal Performance Factors}
\label{sec-principal-performance-factors}

\begin{table}
\caption{Principal Performance Factors}
\label{tab-ppf}
\begin{tabular}{@{}lL{0.39\textwidth}}
\toprule
Abbrv. & Performance Factor \\ \midrule
W       & Workload: resource usage patterns\\
H     & Heterogeneity: type of resources \\
S      & Scale: amount of resources and problem size\\ 
C     & Contention: performance degradation caused by resource sharing \\ \bottomrule
\end{tabular}
\end{table}

Distributed systems are complex and exhibit multiple effects which may cause variations in the performance metrics of jobs or tasks. Methods for predicting performance metrics can be classified by the concrete set of properties of such systems which they take into account, either explicitly or implicitly. The four most influential performance factors are resource usage patterns, resource heterogeneity, scale, and resource contention.

\draft{Resource Usage Pattern}
The first and most fundamental factor is the \emph{resource usage pattern} of the job or task that is to be scheduled. Different jobs or tasks might stress different subsystems of the system: A typical distinction is between compute-intensive jobs, demanding mostly CPU cycles, and data-intensive jobs, requiring mostly main memory and fast data transfer~\cite{Wolski:1999hw,Downey:1997dk,Gibbons:1997em,Delimitrou:2014hz,Miu:2012ix}. In the black-box scenario we consider in this survey, one has limited insight into what a program does, but one may observe its resource usage over time and its interaction with the system. The challenge is that the behavior of a program, and thus its resource usage patterns, can depend in complex ways on the input data and the program arguments.
For example, the execution duration of a scientific simulation might be dominated by the number of scenarios considered, which could be specified as application parameters. 

\draft{Heterogeneity} 
A second important performance factor is \emph{resource heterogeneity} in the target system, denoting varying amounts or kinds of resources across compute nodes. For instance, running a given task on a CPU with higher clock rate is expected to be faster than on a node with a slower CPUs; however, the effect depends much on the resource usage pattern of the task, i.e., whether it is compute-intensive or not. Resource heterogeneity is present in many real-world systems, often induced by extensions of systems over time where each extension utilizes the best machines affordable at the time of the expansion. Resource heterogeneity can also be present by design to allow systems to offer different levels of service~\cite{Ahmad:2012di}. Schedulers critically depend on information about resource heterogeneity to prevent load imbalance~\cite{Cheng:2014dd,Bux:2015wk}. Performance prediction in heterogeneous environments ideally provides a means to normalize observations with respect to hardware, such that predictions for other kinds of hardware can be extrapolated from prior performance observations. 

\draft{Amount of Resources} A third performance factor is the \emph{scale of the problem}. This includes both the amount of resources assigned to an application, \eg, the number of nodes, and the size of the problem, \eg, the amount of input data. The scaling behavior of an application is affected by various factors, such as communication patterns and contention or degree of parallelism. Since the execution times of jobs and tasks typically do not scale linearly with node (or core) count, predicting scale-out behavior is important to sensibly trade off resource demands and application performance~\cite{Barnes:2008ij}. Requesting as few resources as possible is both beneficial for optimizing resource usage, since it avoids idle resources, and for the application itself, since in space-sharing systems queue times can be expected to increase quickly with the amount of requested resources~\cite{Barnes:2008ij, Downey:1997dk}.

\draft{Resource Contention} A fourth performance factor is \emph{resource contention} arising from simultaneous resource usage of different jobs or tasks.
Resources in a distributed environment are often shared and the load imposed by other applications can affect the performance of a task.
For instance, given the performance of a task on an idle node, the performance on a loaded node can be predicted using time series forecasting methods~\cite{Dinda:2001ea}.
Another approach to predict performance degradation due to resource sharing is based on the current load on shared CPU resources, as revealed by hardware performance counters~\cite{Koh:2007jn,Zhao:2016bv}.

\subsection{Prediction Method}
\begin{table}[tbp]
\caption{Prediction Methods}
\label{tab-prediction-method}
\begin{tabular}{@{}lL{.39\textwidth}}
\toprule
Abbrv. & Prediction Method \\ \midrule
CL      & Classification  \\
LL      & Local Learning  \\
R      & Regression    \\
TS       & Time Series    \\ \bottomrule
\end{tabular}
\end{table}

PPM methods predict a certain performance metric for a job or a task based on information about certain performance factors, typically extracted from log files of previous runs. In this section, we summarize the four most common techniques for actually computing predictions, which are classification, local learning, regression, and time series analysis. Note that a detailed description of the technical basis of each of these methods are beyond the scope of our work; the interested readers might refer to~\cite{Hastie:2009wp} for an introduction in classification, regression and local learning, and to~\cite{Wei:2006wh} for time series analysis.

In the following, we focus on application possibilities of these methods to PPM and their specific advantages. 
Regarding the disadvantages, three apparent issues in the context of PPM are
(1) since observations correspond to task or job executions, training data is potentially very expensive to collect and may only be sparsely available.
(2) System workload is difficult to model and its characteristics are likely to change over time~\cite{Downey:1999jv}, such that a trained model might become obsolete.
(3) It can be harder to explain insufficient prediction accuracy and find ways to improve it than for analytical or simulation models.

\draft{Classification}
\emph{Classification} refers to techniques which assign one of a (small) set of classes to a problem instance at hand. Applied to performance prediction, these classes essentially are discretized ranges of the performance metric addressed, and the problem instance is the triple of the distributed system at hand, the job or task to be executed, and the input data. Thus, approaches based on classification are fundamentally based on the assumption that tasks or jobs can be partitioned into categories with similar behavior regarding the performance metric. Predictions can be made based on historical runs, textual job descriptions, or properties of the user or organization submitting a job~\cite{Delimitrou:2014hz,Yang:2005cc, Smith:2004gh,Tsafrir:2007hg,Gibbons:1997em}. Classes may be derived data-driven by clustering historical runs~\cite{Devarakonda:1989wl} or may be specified manually based on expert knowledge. One advantage of the classification approach is that it is relatively simple and has been shown to work reasonably well using only easily collectable job and task metadata~\cite{Gibbons:1997em,Smith:2004gh}. A trained classification model can also provide insights into the properties of a system's workload. 

\draft{Local Learning} 
\emph{Local learning}\footnote{also referred to as instance-based learning, non-parametric regression, and nearest neighbor regression} approaches predict the performance of a job or task to be the same as that of the most similar job or task they have seen in the past. In a more general setting, they also might interpolate between the most similar jobs or tasks to predict the performance metric of interest. Local learning requires a function to compute the similarity of jobs and is based on the hypothesis that jobs similar under this function exhibit similar performance metrics. In contrast to classification approaches, they do not require the definition of distinct classes of jobs. Local approaches are more flexible than classification because they do not require performance metrics to be consistent within classes, and the model can be adapted without retraining by adding new instances or discarding older ones~\cite{Kapadia:1999ug}.

\draft{Regression} 
\emph{Regression} denotes a broad class of methods which map observations into a numerical space and then fit a function over the observations which approximates a given target variable. 
In our setting, the observations are performance factors of the job or task to execute, and the target variable is the performance metric to predict. 
Regression methods usually treat performance factors as numerical variables and consider only a limited set of functions; the by-far most popular methods express a target variable as linear combination of the independent variables. Thus, a trained regression model can also provide explicit insights about what performance factors affect a metric of interest most severely.

\draft{Time Series Analysis}
All previous methods are based on the assumption that the observations they are based on (for classifying jobs, for performing regression, for comparing jobs) are independent from each other. In contrast, \emph{time series analysis} focusses on the temporal development of observations, i.e., they generate predictions based on performance trends and patterns over time. Thus, the temporal order of observations is the primary source of information, assuming that the past development of performance factors dominates future developments. 
These models lend naturally to performance prediction in systems whose resource usage change over time, which for the other methods requires additional steps, such as retraining the model. Another advantage is that performance variation from any source can be captured, as long as the effect is exposed in a predictable pattern over time. 

\section{Task Performance Models}
\label{sec-task-models}

In this section, we survey concrete PPM methods at the task level, that is, performance prediction for single programs running on a single node. 
Such methods are important for schedulers as they allow taking into account the performance variation resulting from placing a task at different compute nodes, or the performance degradation resulting from sharing resources with another task.
The presentation of the approaches is structured according the principal performance factors, as summarized in Table~\ref{tab-task-models}. An approach may occur in multiple of the following subsections, where we consider a paper's contribution relevant to the overall topic of modeling that specific performance factor.

\begin{table*}
\caption{Task Performance Models. The principal performance factors (W,H,S,C) and methods (Mthds.) are abbreviated according to Tables~\ref{tab-ppf} and~\ref{tab-prediction-method}.}
\label{tab-task-models}
\def\arraystretch{1.45}
\begin{tabular}{@{}
L{2cm}L{4.25cm}
l
L{4.2cm}L{2.8cm}l
@{}}
\toprule
References & Context & W H S C & Input & Output & Mthds. \\ \midrule
Marin et al. \cite{Marin:2004hz} & Explore interactions between an application and a platform & ~$\bullet$~~$\bullet$~~$\bullet$~~$\cdot$ & CPU speed and cache sizes, small input instances, application binaries & Execution duration, L1, L2 and TLB cache miss counts & R \\
Hoste et al. \cite{Hoste:2006ji} & Select platform that yields best performance & ~$\bullet$~~$\bullet$~~$\cdot$~~$\cdot$ & Binary instrumentation results, benchmark scores & Execution duration (relative) & LL \\
Yadwadkar et al. \cite{Yadwadkar:2017ida} & Cloud instance selection & ~$\bullet$~~$\bullet$~~$\cdot$~~$\cdot$ & Benchmark results of VM instances, representative task test run & Execution duration & R \\
Ferreira da Silva et al. \cite{daSilva:2015kv} & Predict resource consumption of workflow tasks & ~~~~~~~ & Size of the input files & Execution duration, peak memory usage, disk usage & R \\
Chatzopoulos et al. \cite{Chatzopoulos:2017ja} & Assess scalability, detect bottlenecks & ~$\bullet$~~$\cdot$~~$\bullet$~~$\cdot$ & Sampled durations and stalled cycles for different core counts & Execution duration & R \\
Govindan et al. \cite{Govindan:2011ip} & Virtual machine consolidation & ~$\bullet$~~$\cdot$~~$\cdot$~~$\bullet$ & Memory access pattern, degradation table & Execution duration relative to execution on an idle node & LL \\
Gao et al. \cite{Gao2005} & Effective scheduling in heterogeneous grids & ~$\bullet$~~$\cdot$~~$\cdot$~~$\bullet$ & Execution duration, predicted execution duration of other running tasks & Execution duration & TS \\
Devarakonda and Iyer \cite{Devarakonda1989} & Load balancing in distributed systems & ~$\bullet$~~$\cdot$~~$\cdot$~~$\cdot$ & Execution duration, file I/O, peak memory & Same as input & TS \\
Dinda\cite{Dinda:2001ea,Dinda:dc} & Deadline-constrained scheduling in shared distributed systems & ~$\bullet$~~$\cdot$~~$\cdot$~~$\cdot$ & CPU load, nominal execution duration & Execution duration & TS \\
Zhang et al.\cite{Zhang:2008eg} & Scheduling on distributed infrastructures & ~$\bullet$~~$\cdot$~~$\cdot$~~$\cdot$ & CPU load, nominal execution duration & Execution duration & TS \\
Wolski et al.\cite{Wolski:1999hw,Wolski:1998jq,Wolski:2000hy} & Dynamic scheduling in shared distributed systems & ~$\bullet$~~$\cdot$~~$\cdot$~~$\cdot$ & CPU load, network throughput & Same as input & TS \\
Yang et al.\cite{Yang:2003dj,Yang:2003gv} & Adaptive scheduling on multi-user grids & ~$\bullet$~~$\cdot$~~$\cdot$~~$\cdot$ & CPU load & Same as input & TS \\
Matsunaga and Fortes \cite{Matsunaga:2010kr} & Supporting job schedulers in a heterogeneous cloud or grid environment & ~$\cdot$~~$\bullet$~~$\bullet$~~$\bullet$ & Hardware specifications and benchmark results, input size & Execution duration, resident set size, size of output produced & R \\
Iverson et al. \cite{Iverson:1999cu} & DAG scheduling in heterogeneous cluster & ~$\cdot$~~$\bullet$~~$\bullet$~~$\cdot$ & Numerical description of the input, benchmark results & Execution duration & LL \\
Zhao et al. \cite{Zhao:2016bv} & Improve resource utilization while ``providing QoS guarantees" & ~$\cdot$~~$\cdot$~~$\bullet$~~$\bullet$ & Benchmark results & Execution duration (relative) & R \\ \bottomrule
\end{tabular}
\end{table*}

\subsection{Resource Usage Pattern Modeling at the Task Level}
\label{sec-task-resource-usage-patterns}
Task resource usage patterns can be modeled in different ways.
One way is to characterize them implicitly by observing similarities in behavior and performance of programs. 
This leads to classification or local learning methods based on reusing performance observations from similar tasks. 
Another approach is to model resource usage across repeated executions of the same task.
This leads to time series methods that predict the performance of a task based on the last executions of a task.
Resource usage patterns can also be captured in a fine-grained manner by instrumenting a program or by monitoring hardware performance counters of a CPU. 
Such methods are also used to predict the performance degradation of tasks due to resource contention, which is covered separately in Section~\ref{sec-task-contention}.


One of the earliest time series approaches to PPM was proposed by Devarakonda and Iyer~\cite{Devarakonda:1989wl}.
In their approach, the performance observations are represented as points in a three-dimensional space comprising the dimensions execution duration, peak memory usage, and file I/O.
Using the \textit{k}-means algorithm, clusters of observations are determined.
Subsequently, a Markov model is assembled where each state corresponds to one cluster and transition probabilities are derived from the order of observations.
Based on the previous performance of a task, the performance of the next invocation of a program is predicted, by averaging over cluster centers according to transition probability.

Marin and Mellor-Crummey~\cite{Marin:2004hz} instrument the binary of a task to observe its memory access patterns. 
These patterns are quantified using memory reuse distance histograms, which are based on ``the number of unique memory locations accessed between a pair of accesses to a particular data item''.
The program to predict is executed on small input instances of different sizes, and the bins of the resulting histograms are subjected to regression to describe how the memory access patterns change with input size.

Hoste et al.~\cite{Hoste:2006ji} use a local learning approach to predict execution durations.
They introduce the concept of benchmark space to describe application behavior.
The coordinates of an application are defined by 47 characteristics, such as instruction mix and branch predictability, which are independent of the CPU's microarchitecture (\eg, its cache size and branch predictor size).
These characteristics are obtained for a variety of representative tasks by means of binary instrumentation. 
A new task is profiled in the same way as the representative tasks before its execution to determine its coordinates in benchmark space and thus to quantify its behavioral similarity to the representative tasks. 
Performance measurements of benchmarks that are close to the task in benchmark space are used to predict the performance of the task on different nodes.
Like Marin et al.~\cite{Marin:2004hz}, the authors focus on predicting the best suited node rather than absolute performance values.
Thus, they evaluate their method by comparing predicted and actual task performance rankings of nodes rather than the differences between predicted and actual execution duration.

Zhang et al.~\cite{Zhang:2008eg} developed a method to provide a continuous prediction of CPU load, in contrast to common time-series-based approaches which are limited to a one- or a multiple-step-ahead forecast.
They propose to fit a polynomial curve of second or third order on the CPU load time series using least squares regression.
While this method is able to capture long-term developments in CPU load, it is slow to react to sudden turning points.
To address this issue, the authors propose a means of predicting turning points by comparing the latest CPU load tendencies (\ie, differences between current subsequent measurements) against patterns that occurred in the past and that were associated with a turning point.
If such a pattern can be found, an imminent turning point is deemed likely and the forecast is derived from the historical pattern as opposed to the polynomial curve. 

A seminal work for PPM is the Network Weather Service (NWS)~\cite{Wolski:1999hw} developed by Wolski et al.
It is one of the pioneering works in distributed resource usage monitoring and prediction, which lays the foundation for PPM at higher levels, \eg, to guide scheduling decisions.
The NWS is a distributed performance forecasting framework that continuously measures and forecasts CPU and network performance.
It uses various time series analysis methods~\cite{Wolski:1998jq}, which are applied simultaneously.
The method which has exhibited the smallest prediction error so far is used for the next  forecast.
To keep the intrusiveness of the system low, the frequency at which new forecasts are generated is adjusted automatically, based on the accuracy of earlier forecasts.
The provided bandwidth measurements and forecasts have been extensively used in other performance prediction approaches.
Forecasts of CPU load capture the aggregated short-term resource usage patterns of all tasks running on a node, which can be used to estimate the execution time of an additional task, taking into account the background load.
Refer to Section~\ref{sec-task-contention} for details on approaches that focus on the resource contention performance factor.
To minimize prediction errors when forecasting grid performance, Wu et al.~suggested to extend the autoregressive estimator by applying Kalman and Savitzky-Golay filters to measurement data to reduce noise~\cite{Wu:2010ia}. 
Furthermore, they propose to adaptively adjust the amount of considered measurements to minimize the prediction error, similar to the adaptive sliding window estimator in the NWS.

\subsection{Heterogeneity Modeling at the Task Level}
\label{sec-task-heterogeneity}

Performance variation from hardware heterogeneity arises in various scenarios. 
In a compute cluster, older hardware might be operated side-by-side with newer hardware.
In a cloud computing scenario, customers choose from various instance types and in a virtualized setup, users create virtual machines with custom resource limits.
Predicting the performance of a task on different node types is useful for a scheduler to avoid load imbalance or to prioritize tasks.
In contrast to Section~\ref{sec-task-resource-usage-patterns}, this section focuses on modeling the effects of heterogeneity in the compute infrastructure, \ie, the compute nodes, rather than the workload.

Iverson et al.~\cite{Iverson:1999cu} propose a local learning approach to predict the execution duration of a task on a given node, based on the input size and the hardware characteristics, expressed as benchmark scores.
To be able to reuse observations made on one node for similar nodes, the concept of machine space is introduced.
A node's benchmark scores define its coordinates in machine space and performance observations from nodes that are close in machine space are considered reusable across these nodes.
To avoid high dimensionality in the machine space, a distance-preserving dimensionality reduction is applied.
To decrease the influence of extreme data points, a constant fraction of the largest and smallest observations is discarded. 
The work generally assumes that resources are space-shared, \ie, there is no background load on the nodes. 

Yadwadkar et al.~\cite{Yadwadkar:2017ida} predict the execution duration of a task as a function of the chosen instance types in a cloud computing scenario. Their method is based on an offline profiling phase in which, for each instance type, the performance of a set of benchmarks is measured. In addition to the measured mean and tail performance, the task's load on the CPU, memory, network, disk, and operating system is measured. The combination of a task's performance and resource usage is termed its ``fingerprint''. The collected profiling data is used to train a random forest that learns a mapping from a benchmark's fingerprint on a pair of instance types to a fingerprint on another instance type. In the online phase, users submit a ``representative'' task which is executed and profiled on a pair of instance types to then extrapolate performance on all instance types and deliver recommendations on performance-cost tradeoffs. Although the method works well given a representative task, it does not account for varying input sizes or scaling behavior (problem size) or resource contention.

Marin and Mellor-Crummey~\cite{Marin:2004hz} use resource usage patterns, specifically memory access patterns, to predict the relative performance of a task on different node types.
Given the cache size of a target architecture and the input size, the cache miss counts are predicted.
Similar techniques are used to predict the cost of the computation, which can be combined with the memory hierarchy latency to predict execution duration.
Since exact prediction of execution duration is difficult, the authors focus on ranking the compute nodes according to their suitability for a given task.

Matsunaga and Fortes~\cite{Matsunaga:2010kr} build a regression tree for each program based on input size and hardware characteristics, \eg, CPU speed, cache size, amount of memory, and disk benchmark scores. 
Furthermore, the location of the input data (\eg, on the node or in a network file system) is considered.
Based on these data, they predict execution duration, output file size, and resident set size of the task.
They report that in a few situations where resource performance is non-linear, \eg, a network file system under load, local learning or support vector machines outperform the regression tree, which overall still performs best.
Note that output file sizes and memory consumption are seldomly considered in the literature, Devarakonda and Iyer's early work~\cite{Devarakonda:1989wl} being one of the few examples.
Ferreira da Silva et al.~\cite{daSilva:2015kv} take a similar approach, but use a density-based clustering to partition the observations.
For each cluster of observations, the pearson correlation coefficient is computed to test on a linear relationship. 
If there is one, execution duration, peak memory usage, and disk usage are predicted by regressing on the input file size, otherwise the mean observed resource usage is returned.

Venkataraman et al.~\cite{Venkataraman:2016ww} consider heterogeneity across instance types in a cloud scenario.
They compare the execution duration of several tasks across instance types and use the insights to learn an task-specific regression model that relates execution duration to the input size and the number of nodes.
A separate model has to be learned for each instance type. 
Before investing the cost of training a model for each instance type, they can optimize the costs for a given deadline or budget by choosing the appropriate instance type.

\subsection{Scale Modeling at the Task Level}

As described in Section~\ref{sec-principal-performance-factors}, we consider as the scale factor both to the number of CPU cores assigned to a task and the amount of input data.
The former is important in the context of malleable tasks, \ie, in cases where more CPU cores can be assigned to reduce execution duration. Considering input sizes is important for resource consumption prediction, such as execution duration and peak memory usage.

Zhao et al.~\cite{Zhao:2016bv} predict the execution duration of a task in relationship to the number of threads by applying a thread resource contention model (Section~\ref{sec-task-contention}).
Each thread is treated like an independent single-threaded instance of the task, contending for CPU resources. 
The model takes into account additional CPU cycles incurred by resource contention and thread synchronization, as measured via profiling. 
The predicted total work is divided by the number of threads to obtain an execution duration estimate.
They observe that inter-thread contention for resources dominates the performance variation, whereas thread creation costs and data sharing benefits are negligible.
However, contention effects were only noticeable on a CPU with a relatively small cache, whereas on a CPU with eight cores and 24 MB shared cache, the PARSEC benchmarks~\cite{Bienia:2011wy} exhibited only a few percent of performance degradation due to inter-thread contention.

Chatzopoulos et al.~\cite{Chatzopoulos:2017ja} predict task execution duration as a function of the number of cores used. They use hardware performance counters to measure the number of wasted (``stalled'') CPU cycles due to various reasons, such as waiting for the cache to fetch a line. The key insights are that stalled cycles correlate highly with execution duration, but exhibit scaling trends even before they affect measured runtime. This allows better extrapolation compared to extrapolating execution durations. To fit a model, runtimes and stalled cycles are measured on different numbers of cores and then extrapolated by selecting the best fit from selected rational, cubic, and polynomial functions. When fitting the function, more weight is given to the measurements at higher numbers of cores. The method works well for scale-out predictions on similar platforms but has its limitations when it comes to different architectures, \eg, due to other performance counters or non-uniform memory access. 


Matsunaga and Fortes~\cite{Matsunaga:2010kr} study both the effects of varying input sizes and numbers of threads for two bioinformatics tasks.
They first collect training data by running the tasks on various input sizes and thread counts and then compare several machine learning algorithms in terms of prediction accuracy.
They find that the PQR2 regression tree yields the lowest average percentage error and is also able to capture non-linear effects of execution duration with regard to input size and the number of CPU cores.



\subsection{Contention Modeling at the Task Level}
\label{sec-task-contention}

Due to the trend of increasing numbers of cores per CPU, contention between tasks for shared resources has received much attention.
Early approaches are based on operating-system-level measurement of resource load and task resource demand.
Binary instrumentation can be used to measure task memory access patterns, which can be used to predict performance degradation due to contention for CPU caches. 
Another approach is to profile the tasks' sensitivity for contention to predict the aggregate memory pressure and the according performance degradation for co-running tasks.

Dinda~\cite{Dinda:2001ea} uses autoregressive time series models to predict background load, \ie, CPU utilization by other processes, on a node. 
Load predictions are combined with a task's nominal execution duration, measured on an idle machine, to predict the increase in execution duration on the loaded node. 
The predictions have been used to determine suitable hosts for executing CPU-bound tasks in a distributed system of homogeneous nodes~\cite{Dinda:dc}.
A solution for predicting nominal execution durations, however, is not proposed. 

Yang et al.~\cite{Yang:2003dj} assume the same notion of background load like Dinda~\cite{Dinda:2001ea}.
They also take into account the variation of the background load.
To assign a task, the scheduler favors nodes that have both low background load and expose little variation in background load.
Both quantities can be observed and fed into standard time series predictors, but the authors observe that standard prediction scheme do not sufficiently emphasize the latest few measurements.
Hence, they propose several simple strategies based on the current tendency, \ie,~the difference between the latest and second-to-latest measurements. 
The authors found that their strategies outperform not only the other proposed strategies, but also the predictions of the Network Weather Service. 

Gao et al.~\cite{Gao2005} estimate execution duration based on previous invocations and on the amount of tasks that are already running on a node.
Starting from a scenario where only one program is executed, \emph{node curves} are derived that relate the average execution duration of a task to the number of tasks already running on the node. 
This model is then extended to an arbitrary number of task types. 
The determined execution duration estimates are employed by a sampling-based scheduler, which leaves room for \mbox{(re-)exploring} node performance, while also exploiting favorable task-machine assignments.


Govindan et al.~\cite{Govindan:2011ip} predict the increase in execution duration due to CPU cache contention. 
A program that occupies cache space and bandwidth, the synthetic cache loader, is used to induce synthetic pressure on the CPU's memory subsystem. 
The synthetic cache loader can be configured to occupy a given number of sets and ways in a set-associative cache.
It is assumed that a limited number (\eg, 256) of synthetic cache loader configurations approximately covers all possible memory access patterns.
In a classification-like manner, each task is mapped to the most similar cache loader configuration, for which degradation behavior is known.
Once for every node type, performance degradation of all pairwise cache loader co-locations is measured.
To extend to more than two co-runners, a reduction scheme is proposed that maps the pressure of two co-runners to a single, more aggressive configuration of the cache loader.
This reduction is applied recursively to predict performance degradation with up to four co-runners in the experiments.

Zhao et al.~\cite{Zhao:2016bv} also follow a memory-centric approach using the same basic aggregate pressure approach as in~\cite{Mars:2011un,Govindan:2011ip}.
They distinguish between CPU cache space consumption and cache bandwidth consumption.
This is motivated by the observation that performance degradation can be modeled as a piecewise linear function of the aggregate pressure on both cache space and cache bandwidth.
To measure the cache space and bandwidth consumption, hardware performance counters are used.
To find the piecewise linear function that best models the execution duration of a task, all models in a set of user-defined models are evaluated. 
To reduce the costs of trying all possible model forms, a task-independent model is first created that optimizes the functional form, whereas task-specific parameters for that model are fitted in a second step.


\section{Job Performance Models}
\label{sec-job-models}

In this section, we survey PPM approaches at the job level, that is, prediction methods for complex workloads utilizing several nodes connected through a network. 
Like for tasks, we structure the review of performance models along the principal performance factors from Section~\ref{sec-principal-performance-factors}.

\begin{table*}
\caption{Job Performance Models. The principal performance factors (W,H,S,C) and methods (Mthds.) are abbreviated according to Tables~\ref{tab-ppf} and~\ref{tab-prediction-method}.}
\label{tab-job-models1}
\def\arraystretch{1.45}
\begin{tabular}{@{}
L{2cm}L{4.25cm}
l
L{4.2cm}L{2.8cm}l
@{}}
\toprule
References & Context & W H S C & Input & Output & Mthds. \\ \midrule
Delimitrou et al. \cite{Delimitrou:2014hz} & Data center scheduling & ~$\bullet$~~$\bullet$~~$\bullet$~~$\bullet$ & Short test runs of the application & Performance metrics under different resource configurations & CL \\
Smith \cite{Smith:2007gb} & Site selection in a compute grid & ~$\bullet$~~$\cdot$~~$\bullet$~~$\bullet$ & Job metadata, requested processors, maximum run time & Execution duration, queue time, transfer duration & LL \\
Arslan et al. \cite{Arslan:2016up} & Optimizing end-to-end data transfer parameters for a set of files & ~$\bullet$~~$\cdot$~~$\bullet$~~$\bullet$ & Historical data transfer times, parameters, and background traffic probe data & Transfer duration & R, LL \\
Venkataraman et al. \cite{Venkataraman:2016ww} & Cloud resource configuration & ~$\bullet$~~$\cdot$~~$\bullet$~~$\cdot$ & Input size, number of machines & Execution duration & R \\
Pumma et al. \cite{Pumma:2017gl} & Scientific computing & ~$\bullet$~~$\cdot$~~$\bullet$~~$\cdot$ & Input size, hardware usage during probe execution & Execution duration & CR \\
Alipourfard et al. \cite{Alipourfard:2017tx} & Cloud resource configuration & ~$\cdot$~~$\bullet$~~$\bullet$~~$\cdot$ & Job and initial cloud configuration & Optimized Configuration & R \\
Gaussier et al. \cite{Gaussier:2015fu} & Provide runtime estimates for a backfilling scheduler & ~$\bullet$~~$\cdot$~~$\cdot$~~$\bullet$ & User's job history, job submission time & Execution duration & R \\
Lee et al. \cite{Lee:2007tv} & Tune application parameters for a platform & ~$\bullet$~~$\cdot$~~$\cdot$~~$\cdot$ & Application parameters, sampled execution runs & Execution duration & R \\
Dobber et al. \cite{Dobber2007} & Increase robustness of applications in the grid & ~$\bullet$~~$\cdot$~~$\cdot$~~$\cdot$ & Previous job durations & Execution duration & TS \\
Tsafrir et al. \cite{Tsafrir:2007hg} & Backfilling in a HPC context & ~$\bullet$~~$\cdot$~~$\cdot$~~$\cdot$ & History of user estimated and observed runtimes & Execution duration & TS \\
Sanjay and Vadhiyar \cite{Sanjay:2008ff} & Site selection in a grid environment & ~$\cdot$~~$\bullet$~~$\bullet$~~$\bullet$ & CPU and network benchmark results & Execution duration & R \\
Cunha et al. \cite{Cunha:2017in} & Site selection (cloud or private cluster) & ~$\cdot$~~$\bullet$~~$\bullet$~~$\bullet$ & Metadata, requested Processors, queue state, submission time & Execution duration, queue time & LL \\
Pfeiffer and Wright \cite{Pfeiffer:2008jy} & ``high-performance (HPC) system acquisitions'' & ~$\cdot$~~$\bullet$~~$\cdot$~~$\cdot$ & CPU and network characteristics, HPCC benchmark results & Execution duration & R \\
Li et al. \cite{Li:2007by} & Site selection (grid) & ~$\cdot$~~$\cdot$~~$\bullet$~~$\bullet$ & Metadata, executable name, requested processors, maximum run time, submission time, queue state & Execution duration, queue time & LL \\ \bottomrule
\end{tabular}
\end{table*}

\subsection{Resource Usage Pattern Modeling at the Job Level}
\label{sec-job-pattern}

Compared to the task level, resource usage patterns at the job level include communication between nodes.
In addition, a job resource usage pattern depends on the resource usage patterns of its tasks, leading to potentially much more complex patterns.
Job resource usage patterns can sometimes be inferred from job metadata, such as the name of the application or the submitting user.
The assumption is that jobs with similar metadata will have similar resource usage patterns and thus similar performance metrics.
Historically, analyses of supercomputer workloads showed that categorizing jobs according to their metadata reduces the variance among the observed runtimes~\cite{Feitelson:1995jj}.
Not very surprisingly, invocations of the same job, submitted by the same user on the same number of requested processors have been observed to expose much less runtime variation than the set of all jobs observed in a cluster.
Gibbons~\cite{Gibbons:1997em} verified this observation and used it to predict job execution duration by simply averaging runtimes of prior jobs with the same metadata.
 Downey~\cite{Downey:1997dk} makes a similar observation in a San Diego Supercomputer Center workload, where job runtimes approximately followed a log-uniform distribution, with different parameters for the queues for short, medium, and long jobs.

\addtocounter{table}{-1}
\begin{table*}
\caption{(continued) Job Performance Models.}
\label{tab-job-models3}
\def\arraystretch{1.45}
\begin{tabular}{@{}
L{2cm}L{4.25cm}
l
L{4.2cm}L{2.8cm}l
@{}}
\toprule
References & Context & W H S C & Input & Output & Mthds. \\ \midrule
Downey \cite{Downey:1997dk} & Site selection & ~$\cdot$~~$\cdot$~~$\bullet$~~$\bullet$ & Number of running jobs, their age, and occupied number of processors, requested number of processors & Queue time & R \\
Barnes et al.  \cite{Barnes:2008ij} & Improve cluster efficiency & ~$\cdot$~~$\cdot$~~$\bullet$~~$\cdot$ & Application parameters, number of processors & Execution duration & R \\
Gibbons \cite{Gibbons:1997em} & Provide estimates to scheduler & ~$\cdot$~~$\cdot$~~$\bullet$~~$\cdot$ & Executable name, user name, assigned number of processors, current job age & Execution duration & CL, R \\
Wu et al. \cite{Wu:2010ia} & Scheduling and load balancing on the grid & ~$\cdot$~~$\cdot$~~$\cdot$~~$\bullet$ & CPU load history & CPU load & TS \\
Brevik and Nurmi \cite{Brevik:2006hv,Nurmi:2007vt} & Estimating queueing delays for batch jobs & ~$\cdot$~~$\cdot$~~$\cdot$~~$\bullet$ & Queue time history & Queue time & TS \\ \bottomrule
\end{tabular}
\end{table*}

\draft{Smith and followups: the template method}
Smith et al.~\cite{Smith:1999jf} build on the observations by Gibbons~\cite{Gibbons:1997em} to predict job execution durations from scheduler logs.
They coined the term \emph{template} for a set of metadata attributes used to partition the observations into groups.
The execution duration for a new job is predicted by looking up prior observations in the according category and aggregating them either by using the arithmetic mean or by regression over the number of assigned processors.
A genetic algorithm is used to determine the metadata features to group by and the aggregation method.
Instead of searching for a single best template, they also propose to search for ``template sets'' that consist of one to ten templates, each giving one prediction with a confidence interval for a new job. 
Li et al.~\cite{Li:2004iv} found that averaging all templates' predictions can improve prediction accuracy.
The template approach later has been extended by considering more job attributes, \eg, the topology of a workflow-structured application~\cite{Nadeem:2013cf}, predicting other resource types, \eg, memory usage~\cite{Piro:2009kx}, and by using other prediction methods within the categories, \eg, time series predictors~\cite{Li:2004iv,Sonmez:2009cr}.

Tsafrir et al.~\cite{Tsafrir:2007hg} predict the runtime of a job as the average of the runtimes of the last two submitted jobs of the same user.
This is a time series approach that aims more at the resource usage pattern of the user rather than the job.
The authors compare several variations of the above scheme 
and conclude that no method works best for all systems and workloads.
Other time series-based approaches include Sonmez et al.~\cite{Sonmez:2009cr}, who show that the accuracy of time series predictors for job runtime and queue time predictions can be boosted when building separate models for each user, computing site, or both.
Gaussier et al.~\cite{Gaussier:2015fu} propose a set of features similar to the above time series features for a linear regression model.
They include a variety of information about the user's job history, \eg, the execution durations of the last three jobs and the time since the last job of that user was completed.
Their cost function penalizes the underestimation of execution duration stronger than overprediction because backfilling schedulers need to take corrective actions when jobs exceed their reservation, in the worst case even killing the job.

Dobber et al.~\cite{Dobber2007} compare several time series predictors for execution duration prediction. 
The proposed adaptive exponential smoothing technique improves upon the basic exponential smoothing predictor implemented in the Network Weather Service~\cite{Wolski:1999hw} (see Section~\ref{sec-job-interference}) by continuously adjusting the smoothing parameter based on the observed forecast error.
This way, the weight of the latest measurement is increased if, for instance, the last forecast was found to be inaccurate.
They also propose a dynamic exponential smoothing method, which combines a sliding window approach with the adaptive exponential smoothing predictor.
Results were found to be comparable or even preferable to the Network Weather Service, due to an increased robustness to peaks and changes in performance. 

\draft{state of the art classification: regression trees and recommender systems}
Recently, more sophisticated classification methods have been applied to job performance prediction.
The performance predictions of Matsunaga et al.~\cite{Matsunaga:2010kr} and Dwyer et al.~\cite{Dwyer:2012kg} are based on regression trees, a technique that combines classification and regression~\cite{Rokach:2014ep}.
Both predict different metrics, but identify regression trees as most accurate among a wide range of machine learning techniques.
This contradicts the observations by Smith et al.~\cite{Smith:1999jf} and Tsafrir~\cite{Tsafrir:2007hg}, who favor simple predictors like the mean over a regression-based aggregation of observations.

Pumma et al.~\cite{Pumma:2017gl} use an approach that completely ignores metadata. Instead, each job is run for a short while and profiled using platform independent low level features derived from hardware performance counters.
These include for instance instruction mix, branch predictability, and address distances between successive cache accesses. Using a pre-built classification scheme based on seven major types (``Dwarfs'') of scientific applications~\cite{Asanovic:br}, the job is assigned to a job class that implies its resource usage pattern. 
The C4.5 decision tree is trained using manually labelled benchmarks from three benchmark suites~\cite{Che:2009bc,Bailey:1991vs,Kaiser:2011wf}.
For each job class, linear regression is used to model runtime as a function of input size and observed resource usage patterns (the measured low level characteristics).

Delimitrou et al.~\cite{Delimitrou:2014hz} use singular value decomposition to classify an incoming job based on a short test run of the application. 
The idea is to extrapolate the performance information for untested configurations from other applications with similar resource usage patterns, as revealed by the test runs.
The assumption here is again that the workload can be partitioned into jobs that expose similar behavior under the principal performance factors.

Lee et al.~\cite{Lee:2007tv,Ipek:2005fd} analyze the impact of application parameters on the execution duration of two high performance scientific applications.
These applications have large parameter spaces comprising both parameters that define the workload, \eg, working set size per processor, and those that define how to execute the workload, \eg, which broadcast method to use.
Such parameters clearly affect the resource usage pattern, and through the latter parameters, their work gains a performance tuning aspect.
Moreover, a statistical analysis is incorporated, which indicates non-linear relationships between parameters and execution duration.
The authors thus choose neural networks and piecewise polynomial regression for modeling the relationship.


\subsection{Heterogeneity Modeling at the Job Level}

Heterogeneity at the job level is more complex than at the task level, because it may affect different parts of a job.
At the task level, hardware heterogeneity, \ie, hardware specifications affect the entire task.
At the job level, parts of the job may run on different nodes, and these parts depend on the scheduler, its awareness of hardware differences and its ability to incorporate them in its decisions.
A typical problem in grid computing is to choose a site to run a job on, under the premise that resource characteristics of the sites are heterogeneous.
A variant of the site selection problem is computing in a hybrid cloud, in which, for instance, some of the compute resources are owned by the organization and additional resources can be acquired via cloud computing on demand.
The estimation of job finish times depends on predictions of queue time, execution duration, and file transfer duration.\footnote{We address queue time estimation in Section~\ref{sec-job-interference}, as queue times are a result of jobs contending for resources.}

Sanjay and Vadhiyar~\cite{Sanjay:2008ff} model execution duration of a parallel application as the sum of computation duration and non-overlapped communication duration. 
Both durations are modeled by regression, using a mixture of 77 polynomial and logarithmic functions.
They address cross-platform performance prediction by starting from a regression model learned on a reference platform, conducting dedicated small-scale experiments on the reference and target platform and using the observed performance ratios to scale the coefficients of the reference model.

Cunha et al.~\cite{Cunha:2017in} consider performance differences between a local cluster and cloud resources.
They assume that executing in the cloud slows down execution by a constant, application-dependent factor, for which they propose a linear model and an empirical model, based on 
relative performance of eight applications on three clusters and on three cloud platforms.
Cunha et al. acknowledge that it is difficult to predict relative performance because of various factors, such as the network performance.
To alleviate this issue, they use a ``cutoff function'' that favors executing on the local cluster in case of high uncertainty associated with the predictions.

Pfeiffer and Wright~\cite{Pfeiffer:2008jy} assume a homogeneous cluster, but build models that regress on the performance characteristics of a node type and thus have the possibility to predict performance for different (but homogeneous) clusters. 
They also express job execution duration as the sum of computation time and communication time that is not overlapped with computation.
The former is modeled as a linear combination of CPU and memory benchmark results, whereas the latter is expressed as a linear combination of interconnect bandwidth and latency benchmark results.
Both fits are performed using non-negative least squares regression, on a subset of the benchmarks automatically selected by backward elimination~\cite{Draper:105677}.

Alipourfard et al.~\cite{Alipourfard:2017tx} propose a search method to select cloud configurations that optimize the performance of a job. A configuration comprises the number of virtual machine instances, number of cores, CPU speed, RAM per core, disk count and speed, and network capacity. A gaussian process model is used to estimate the uncertainty associated with a configuration and to select the one that has the largest potential for increasing performance. The configuration evaluated and the model is updated until expected improvements fall below a threshold. A limitation of the method is its reliance on representative workloads. Since the search procedure is too slow to be applied to a job at hand, it is applied offline to representative workloads instead. This is particularly problematic with respect to changes in input size, which the method does not model explicitly. This makes the method a perfect fit recurring jobs, which often operate on data sets of similar size.

\subsection{Scale Modeling at the Job Level}
\label{sec-job-scale}

The scale of a job refers both to the problem size, \eg, amount of input data, and the amount of resources assigned to the job.
Compared to tasks, the effects of scale can be much more pronounced and thus require careful decision making. 
For instance, choosing the right amount of resources is important because both undersizing and oversizing can have adverse effects.
Allocating too many resources might result in an inefficient utilization, \eg, resources left idle although reserved, or cause additional parallelization overheads that diminish or even outweigh performance gains.
Also, wait time in a cluster resource queue usually increases with the amount of requested resources.
Too few resources on the other hand may result in a missed deadline or the failure of the job. 
Since scale-out is often modeled as a linear or sub-linear function, regression methods are a natural fit to this problem. 
Another option is to include the amount of resources, \eg, the number of requested cores into the distance metric of a local learning approach~\cite{Smith:2007gb,Li:2007by}.

Barnes et al.~\cite{Barnes:2008ij} regress execution duration both on application parameters and processor count.
Here, the application parameters describe the problem size, \eg, the resolution of a fluid dynamics simulation.
They try both a linear and quadratic form for the processor count and pick the better fit.
All variables are transformed by taking their logarithm before regression to avoid the large execution durations dominating the short ones in term of influencing the model coefficients.
They observe that regressing separately on the time spent for computation and communication works better, as these two quantities tend to scale differently.

Venkataraman et al.~\cite{Venkataraman:2016ww} model application execution duration as a function of input size and the number of nodes. 
They argue that execution duration can be modeled as the sum of constant costs, the costs of parallelizable work, a logarithmic communication cost for aggregating data in a tree pattern, and linear per-machine overheads.
They use a non-negative least squares regression to find the coefficients of these cost factors given some runs of an application at different scales.



Arslan et al.~\cite{Arslan:2016up} predict the duration of large file transfers on the network, focusing on 
three parameters: amount of data, concurrency, and parallelism, where concurrency denotes the transfer of several files in parallel and parallelism refers to transferring the blocks of a single file with several TCP streams in parallel. 
The authors apply local learning to retrieve the most similar observations to the current network load situation as found by historical probe data.
Then, the relationship between the transfer parameters concurrency, and parallelism is found using polynomial regression.
These models are finally used to predict optimal settings for a transfer of a set of files.
A similar approach is proposed by Liu et al.~\cite{Liu:2017bm}, who also consider the two file transfer parameters concurrency and parallelism. 
Based on large Globus GridFTP logs, they engineer additional features, such as the load on the two endpoints.
Finally, they build a regression model that relates the transfer configuration to its duration.

\subsection{Contention Modeling at the Job Level}
\label{sec-job-interference}

Job resource contention commonly arises in two ways: queue times for compute jobs and background load on the network for file transfer jobs.
We cover these scenarios in the following subsections. 

\subsubsection{Queue Time Prediction}

On space-shared clusters, jobs queue up for resources and the scheduler decides which job to run next.
For a metascheduler, \ie, a scheduler that submits a job to one of several sites, 
predictions of queue time are important, as they add up to the job execution duration.
Queue time can be predicted based on the queued jobs, their predicted execution durations, predicted new job arrivals, and the scheduling policy applied to the queue.

Downey~\cite{Downey:1997dk} considers the case of a simple first-come-first-served scheduler to predict wait time for the job at the head of the queue based on the number of processors it has requested. 
He first fits a log-uniform distribution to the execution durations of the jobs in the queues for short, medium, and long jobs.
Given the current age of a job, the probability of running for another $t$ minutes can be estimated.
Using this, the probability of a number of processors becoming free during the next $t$ minutes can be estimated. 
Smith et al.~\cite{Smith:1999jf} applied the template method (see Section~\ref{sec-job-pattern}) they developed for runtime predictions to queue time prediction and report improvements over Downey's method.
They also consider more schedulers, including a backfilling scheduler. 
Li et al.~\cite{Li:2007by} combine the metadata-based classification of Smith and an idea similar to Downey's approach in a local learning approach.
The state of the resource is characterized by attributes like the number of currently running jobs and their estimated runtimes.
A genetic algorithm is used to optimize various hyperparameters.

Brevik and Nurmi et al.~\cite{Brevik:2006hv,Nurmi:2007vt} propose a time series-based approach for queue time prediction.
The first component is a conservative quantile estimator for distributions, which is applied to queue waiting times, disregarding the temporal order of observations.
The second component is a change point detector.
The assumption is that queue times are usually consistent for an extended period of time before a change causes entrance into another consistent phase.
A change point is assumed after observing a fixed number of queue times beyond the .95 quantile of the current distribution of queue times. 
Similar to Smith's template approach, queue time observations are grouped according to their job metadata.


Sonmez et al.~\cite{Sonmez:2009cr} conducted a comparative evaluation of estimators for job execution duration and for queue time on grid job logs.
This includes various time series methods and Nurmi's method~\cite{Nurmi:2007vt}.
They confirm that partitioning observations, \eg, by site, user or both, improves predictions, although they do not observe strong improvements in the scheduling quality in a trace-based simulation.
This contradicts other studies~\cite{Tsafrir:2007hg,Casanova:2000iw}, emphasizing the difficulty of comparing different systems under various workloads.

\subsubsection{File Transfer Duration Prediction}

Resource contention also plays an important role for file transfers.
The replica selection problem arises when multiple copies (replicas) of a file reside in the distributed environment.
A scheduler may want to select the one with shortest transfer duration to a specific endpoint.
Here, resource contention has to be considered in the form of background load on the network and the endpoints.


Faerman et al.~\cite{Faerman:1999iq} address the problem of predicting file transfer durations based on network bandwidth probes. 
The problem is that, to keep probing overhead low, the probes are usually small in comparison to actual files.
Instead of just dividing file size by estimated bandwidth, the authors regress transfer duration on bandwidth probe measurements (a single coefficient regression model), which presumably captures the overhead of larger file transfers. 
Swany and Wolski~\cite{2002swany} also map probed bandwidth values to file transfer durations.
Instead of a regression model, they map the cumulative distribution function of the bandwidth probe values $\text{CDF}_B$ to the cumulative distribution of file transfer durations $\text{CDF}_D$.
The forecast is obtained by looking up $\text{CDF}_D^{-1}$ for $\text{CDF}_B(B)$.
The bandwidth value $B$ is suggested to be obtained from a Network Weather Service forecast.
Vazhkudai and Schopf~\cite{2003vazhkudai} show that also considering disk load values improved predictions. 
A comprehensive study considering disk load and several additional factors was recently conducted by Liu et al.~\cite{Liu:2017bm}.

Qiao et al.~\cite{Qiao:2004jc} consider the impact of the frequency of bandwidth probes.
They claim that for small files, a higher-resolution signal is needed, whereas for large files, a coarse-grained observation history will suffice.
They evaluate numerous network packet traces with bandwidth utilization data captured on WANs and LANs in combination with smoothing techniques to adapt the time scale of the bandwidth utilization time series. 
Subsequently, they apply various prediction methods to this data, including the mean and last value alongside autoregression approaches and nonlinear schemes.
They find that often there is a sweet spot in the resolution of the bandwidth probe signal which improves predictability, but the optimal resolution depends on unidentified criteria. 
Notably, simple models are shown to be competitive with more complex prediction schemes, a result that has already been observed by Vazhkudai and Schopf~\cite{2003vazhkudai}.

\section{Obtaining Performance Indicators}
\label{sec-integration}
\label{sec-monitoring}

\begin{figure}[!t]
\centering
\includegraphics[width=.48\textwidth]{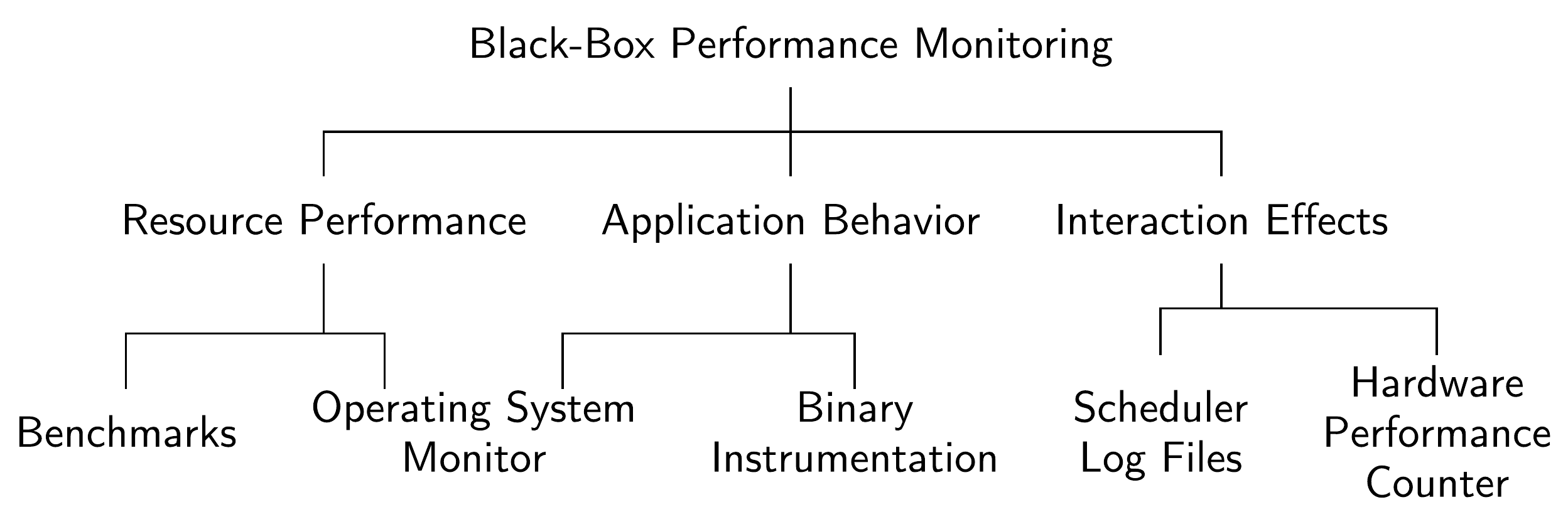}
\caption{Methods for performance monitoring that do not require modification of the workload or execution system.}
\label{fig-monitoring-methods}
\end{figure}

All methods presented in this survey rely on performance measurements of jobs or tasks that do not require modifications to the workload. We thus give a brief overview of common tools and methods for black-box performance monitoring.
To observe performance of jobs and tasks in a non-intrusive way, various tools and methods are available.
These produce data at different levels of granularity and at different levels of overhead.
As shown in Figure~\ref{fig-monitoring-methods}, one can distinguish approaches that aim at measuring the behavior of an application, the performance of the resources, and the interaction between both.\footnote{Another common source of information are application logs, but these require information on how to parse, interpret, and use the relevant information~\cite{Tan:2008wp}, which is not available in a black-box scenario.}
In the following, we briefly explain each approach, give usage examples, cite exemplary tools and discuss their overhead.

\draft{benchmarks and load probes}
\emph{Benchmarks} can be used to measure the performance of a node with respect to some representative task.
Measuring such task-specific performance is important because resource consumption may depend on complex interactions, \eg, between CPU, memory and mass storage.
To compare the performance of two nodes with respect to some task, their benchmark scores are likely more useful than their hardware specifications, given that the task to schedule is similar to the benchmark task. 
Popular benchmark suites are SPEC \cite{Henning:2006bv} and PARSEC~\cite{Bienia:2011wy}.
As benchmarks are usually only executed once, their overhead is negligible. 
\emph{Load probes} are a related technique, which aim at the current, transient performance of a subsystem, usually the network.
A small amount of data is sent in order to measure the current available bandwidth and latency.
This technique was used for instance in the Network Weather Service \cite{Wolski:1999hw} and many others \cite{Yuan:hn,Faerman:1999iq}.
With respect to overhead, probe sizes closer to the actual transfer size are more representative, but also incur higher overhead~\cite{Faerman:1999iq}.

\draft{operating system}
\emph{Operating systems} provide means to inspect the current load on a node's devices and all running processes.
In a virtualized environment, the hypervisor plays a similar role as the operating system and can be used to collect performance measurements~\cite{Koh:2007jn}.
In a Linux environment, the virtual \texttt{proc} file system exposes performance information, \eg, about the CPU and memory usage of individual processes. 
Another option is to monitor the task's interaction with the operating system by intercepting system calls.
This can be used, \eg, to measure the amount of data a task reads and writes, which is an aspect of application behavior, rather than resource performance.
Tools like \texttt{uptime}, \texttt{vmstat}, and \texttt{iostat} are useful to determine the current load of CPU or mass storage devices~\cite{Wolski:1999hw, 2003vazhkudai}.
As an example of a higher-level tool, Kickstart~\cite{Juve:2015ga} can be used to wrap the execution of any task and collect comprehensive performance information.
The overhead of polling operating system performance metrics is low, while intercepting system calls for tasks can be expensive, depending on how much information is collected.


\draft{binary instrumentation}
\emph{Binary instrumentation} is a technique that modifies the executable of a program to collect information about its behavior while it executes.
In general, it is challenging to obtain hardware-independent information about a black-box task, because its observed performance is always a product of the task behavior and the resources it interacts with.
Using binary instrumentation, memory access patterns can be characterized in a microarchitecture independent manner~\cite{Hoste:2007ch}, giving some low-level insights into program behavior.
For instance, Marin et al.~\cite{Marin:2004hz} use binary instrumentation to measure memory access patterns to select a CPU with appropriate cache size.
Typical tools for binary instrumentation are PEBIL~\cite{Laurenzano:2010dx} and Dyninst~\cite{Bernat:2011cg}.
This technique has relatively high overhead and the execution duration of the binary can increase considerably.


\draft{scheduler log files}
\emph{Scheduler log files} are a widely used source for training prediction models.
They usually comprise metadata and performance data, such as the submitting user, the application name and the observed resource consumption of a job.
Since high-level performance metrics like the job execution duration are a joint product of resource usage pattern and hardware characteristics, we classify log files as a monitoring technique for interaction effects.
Comprehensive logs from large-scale systems in science and industry are occasionally published, \eg, the parallel workloads archive~\cite{Feitelson:2014vc} or Google cluster traces~\cite{Reiss:2012bm}.
Each application can also log its own performance data, and a tailored analysis can yield valuable information~\cite{Tan:2008wp}, but since it cannot be exploited in an application-independent way, such approaches are out of the scope for this work.
Since the scheduler usually collects accounting information anyways during operation, log data comes virtually for free. 

\draft{hardware performance counters}
\emph{Hardware performance counters} provide low level insights into task behavior.
Performance counters are dedicated registers in a CPU that can be used for measuring events, such as cache misses, branch mispredictions, or the total number of CPU cycles consumed by a process.
They are often used to reveal the performance degradation due to sharing resources with another task, \eg, by measuring an increase in cache misses or cycles per instruction compared to executing with exclusive access to resources. 
For instance, Zhao et al.~\cite{Zhao:2016bv} and Dwyer et al.~\cite{Dwyer:2012kg} predict the slowdown experienced by two applications contending for shared CPU resources based on performance counter observations.
As the configuration of these hardware performance counters is microarchitecture-specific, typically a high-level interface like PAPI~\cite{2001dongarra} or LIKWID~\cite{Treibig:2010hb} is used for interacting with performance counters.
Because measurement takes place in hardware, hardware performance counters incur only negligible overhead.

\section{Open Issues and Directions for Research}
\label{sec-open-issues}

\begin{table}[]
\centering
\caption{Open topics in PPM research. Benefits are abbreviated as prediction accuracy (A), model generalizability (G), and overhead reduction (O).}
\label{tab-opportunities}
\begin{tabular}{@{}lL{4.3cm}l@{}}
\toprule
Domain & Opportunities & Benefits \\ \midrule
Prediction & Online Learning & A,G,O \\
 & Active Learning & A,O \\
 & Multi-resource models & A \\
 & Temporal resource usage & A \\
 & Multi-level parallelism & A \\
 & Continuous prediction & A \\ 
\midrule
Data collection & Resource-agnostic workload patterns & A,G,O \\
 & Monitoring data integration & A,G \\
 \midrule
Decision making & Holistic approaches & A,G,O \\
 & Sufficient accuracy models & O \\
\bottomrule
\end{tabular}
\end{table}

The reviewed literature demonstrates the long-standing and evolving research in predictive performance modeling. In this section, we identify future trends and research opportunities in the field. 
We tag research opportunities according to three classes of improvements. 
Most research aims at improving the prediction accuracy (A) delivered by a predictor in a certain scenario. A second, often opposing goal is model generalizability~(G), \ie, the ability of a predictor to perform well in a large range of scenarios. Thirdly, collecting training data for predictors can incur significant costs. Therefore, reducing the overhead (O) associated to integrating PPM into a system is another research goal. An overview of the research opportunities discussed in the following is given in Table~\ref{tab-opportunities}.

\subsection{Prediction}
Often, PPM methods involve an offline phase in which performance measurements are taken, models are trained, and hyperparameters are chosen. On the upside, this approach allows for costly collection of high quality training data (since it is usually performed only once) and keeps the complexity of the production system low. On the downside, this approach cannot directly benefit from data collected during system operation. Performance may degrade over time, as the system or its workload changes. Thus, we see a major research opportunity in approaching training and prediction in an \textit{online} fashion, \ie, collecting training data and updating prediction models during system operation.

A major challenge in PPM using machine learning is that training data sets can be expensive to obtain, as each observation might come at significant costs. In addition to using data that is generated during system operation, \textit{active learning}~\cite{Olsson:2009wb} approaches and sampling strategies~\cite{Sarkar:2015fr,Singh:2017gu} might reduce data collection overhead and increase accuracy by achieving better coverage of the feature space.

In the literature, models for predicting execution duration dominate. Recently however, other resource types such as memory~\cite{Rodrigues:2017bi,Reiss:2016tj} shift more into focus. A research opportunity would be to investigate and predict the \textit{interdependent usage of resources} rather than predicting the usage of each resource type separately~\cite{Matsunaga:2010kr,Devarakonda:1989wl}.
Predictions regarding disk and memory usage are essential to more efficiently provision resources, \eg, in a cloud scenario~\cite{Deelman:2012wc}. Predictions for resources other than the CPU gain importance as I/O- and memory-intensive workloads arise~\cite{Pennisi:2011bx} in distributed computing.

A related research direction is the exploration of \textit{temporal variations}. Resource usage of tasks and jobs usually varies across time and predicting the usage of multiple resources as a (multi-dimensional) time series rather than as their aggregate metrics could lead to more sophisticated schemes to improve resource utilization and reduce contention between jobs and tasks. Phase analysis~\cite{Sherwood:2003vr,Zhang:2015gj} is a related research field that explores the resource usage of programs on a cycle-accurate level, but we see a research gap in exploring the use of several resources in a black-box fashion and across various time scales.

Large-scale computing involves \textit{multiple levels of parallelism}, ranging from multi-core architectures over clusters to grid and federated cloud~\cite{Coutinho:2015wy} scenarios. This flexibility offers new trade-offs with severe performance implications, for instance when having to choose between fewer machines with more cores and more machines with fewer cores. Resources like memory, network, and disks per compute node contribute to complex performance behavior~\cite{Alipourfard:2017tx}. Predicting performance for various resource configurations and granularities of parallelization is especially pressing in cloud scenarios, where users have to choose from a large range of compute node types~\cite{Venkataraman:2016ww}.

Another challenge is handling errors from PPM. While all methods strive for high average accuracy, maximum errors can still be high. To mitigate those effects and to avoid catastrophic decisions, schemes to correct decisions~\cite{Mishra:2017dr,Jeon:2016hu} and update predictions~\cite{Delimitrou:2014hz} are emerging. Another possible research direction is to extend single predictions before making a decision with \textit{continuous predictions} that account for newly arriving monitoring information and provide a decision maker with updated, more accurate predictions. 

\subsection{Data Collection}

A largely unexplored aspect is the collection of \textit{resource-agnostic} workload patterns. Observed performance is usually a product of a task's or job's resource usage patterns and a specific resource configuration. For example, the number of cache misses is a product of the application's memory access pattern and the cache size. To predict performance across various resource configurations, it would be helpful to quantify application behavior in a resource-independent way. Approaches that collect such information in a black-box fashion exist~\cite{Hoste:2007ch,Marin:2004hz} but incur high runtime overheads and cover only low-level performance metrics. Detecting and accounting for more general interaction effects between workload and resources would support PPM by reducing the amount of data that needs to be collected and would help to generalize better to other resource configurations.

Various technologies have transformed the landscape of distributed computing during the last years. For instance, the cgroups kernel feature in Linux operating systems now allows for better monitoring and control of the resource usage of processes or groups of processes. On the networking side, software defined networking has introduced new opportunities to monitor and thus predict resource usage. Hardware performance counters allow for low-level characterization of application performance. Aggregating this low-level information to more descriptive metrics and integrating the data of various sources to characterize an application's resource usage is both an engineering and a research challenge.

\subsection{Decision Making}

There is a dichotomy in the literature between scheduling and prediction: scheduling research often takes the availability of performance estimates for granted and research on PPM seldomly explores its impact on scheduling.
We see a large research potential in the exploration of \textit{holistic approaches} to prediction, data collection, and decision making.
For instance, improving prediction quality could be added the objectives of a scheduler. By integrating scheduling and prediction, systems that continuously improve accuracy, adopt to changing workloads, and make optimal use of performance observations during system operation become feasible.

In this context, the question of \textit{sufficient prediction accuracy} arises, \ie, how scheduling quality is affected by prediction accuracy. This relationship has been empirically evaluated before~\cite{Gibbons:1997em,Casanova:2000iw,Tsafrir:2007hg}, but the results are ambiguous and still lack a theoretical foundation.
Further research in this area might enable models of sufficient accuracy, that allow for better trade-offs between data collection efforts, the resulting prediction accuracy, and the returns of improved decision making.

\section{Conclusion}

Predictive performance modeling methods are important to support distributed computing at growing scales and levels of automation, under increasing hardware complexity and workload diversity.
We surveyed approaches that provide performance predictions to improve the utilization, throughput, and other efficiency metrics of distributed computing systems.
The approaches were compared in various dimensions, most importantly the principal performance factors they account for, \ie, resource usage patterns, resource heterogeneity, problem scale, and resource contention. 
From this perspective, we reviewed and classified the literature according to prediction model, input and output data, and use cases.

In conclusion, machine learning methods are an approach to the problem of efficiently managing and planning resources in distributed systems under various factors of uncertainty.
They offer greater versatility and reduced development costs compared to simulation and analytical performance models that are tailor-made for specific applications and infrastructures.
The key idea is to put a stronger emphasis on observed performance factors as opposed to the performance factors anticipated by an analytical model or simulation, which offers greater flexibility at the price of reduced prediction accuracy.
However, we believe that machine learning approaches are necessary to make PPM available in practice for a large range of applications and computing platforms.
Further developments in PPM and their integration into scheduling algorithms promise improved resource utilization, reduced execution times and costs, and more satisfied users.

\section*{Acknowledgements}
Carl Witt, Marc Bux, and Wladislaw Gusew have received funding by the Deutsche Forschungsgemeinschaft (DFG) through the SOAMED graduate school (GRK 1651).

\section*{References}

\bibliography{carl-witt-entire-library,bux,wlad-gusew-entire-library,wlad-gusew-entire-confs}

\end{document}